\documentclass[12pt]{iopart}
\usepackage{iopams}
\usepackage{graphicx}
\usepackage{cite}
\usepackage[english]{babel}

\begin{document}

\title[Oscillating spin-orbit interaction as a source of spin-polarized wave packets]{Oscillating spin-orbit interaction as a source of spin-polarized wave packets in two-terminal nanoscale devices}
\author{Viktor Szaszk\'{o}-Bog\'{a}r$^{1,2}$, P\'{e}ter F\"{o}ldi$^{1}$, F. M. Peeters$^{2}$}
\address{$^{1}$ Department of Theoretical Physics, University of Szeged, Tisza Lajos k\"{o}r%
\'{u}t 84, H-6720 Szeged, Hungary}
\address{$^{2}$ Departement Fysica, Universiteit Antwerpen, Groenenborgerlaan 171, B-2020
Antwerpen, Belgium}

%\ead{vszaszko@physx.u-szeged.hu}

\begin{abstract}
Ballistic transport through nanoscale devices with time-dependent Rashba-type spin-orbit interaction (SOI) can lead to spin-polarized wave packets that appear even for completely unpolarized input. The SOI that oscillates in a finite domain generates density and spin polarization fluctuations that leave the region as propagating waves. Particularly, spin polarization has space and time dependence even in regions without SOI. Our results are based on an analytic solution of the time-dependent Schr\"odinger equation. The relevant Floquet quasi-energies that are obtained appear in the energy spectrum of both the transmitted and reflected waves.
\end{abstract}

\maketitle

\section{Introduction}
The presence of Rashba-type \cite{R60} spin-orbit interaction \cite{NATE97,G00} (SOI) in semiconducting materials can lead to spin-dependent quantum mechanical phenomena that are of both fundamental and practical interest. In rings and loop geometries fundamental interference effects can be observed, and they can have important applications as well, mainly due to the fact that the strength of the SOI has been proven to be experimentally tuneable \cite{NATE97,G00}. The relevance of the physical system motivated
extensive studies\cite%
{BIA84,AHIYE12,NMT99,FR04,MPV04,FMBP05,SzP05,SN05,KMGA05,BO07,CHR07,BO08,CPC08,KFBP08c,M10, DMK11,FE12,NMMG11,NEEA13}
of quantum rings or systems of them.

Since the strength of the SOI -- that determines the spin sensitive behavior of the devices -- can be controlled by external gate voltages, the question what effects appear if these voltages are time dependent arise naturally. This point is to be investigated in the current paper.

Transport phenomena in the presence of an alternating field can be strongly inelastic, like in the case of photon-assisted tunneling \cite{TG63}. Barriers with oscillating height were investigated in a series of papers, mainly in the context of traversal time and photon assisted transport, see e.g.~Refs.~\cite{BM82,X04,WLL06}. Floquet's theory \cite{F883} was proven to be an efficient tool for the investigation of time-dependent transport in various materials \cite{MB04,NSzP12,WC06,LR99,LSS12,AESB12}. Transport related problems with oscillating SOI have been studied in Refs.~\cite{WC07,FBKP09} for a ring, and in Ref.~\cite{RCM08}
for a ring-dot system. Additionally, Ref.~\cite{SS13} demonstrated gate-driven electric dipole spin resonances in a quantum wire subjected to both static and oscillating potentials. These works mainly focused on spin currents and time averaged transport properties; in the following we take a different point of view by presenting time-resolved results.

As we shall see, the oscillation of the spin-orbit interaction strength leads to effects that are qualitatively different from the case of constant SOI. Most remarkably, we point out the existence of time instants when the simple geometries we consider produce spin-polarized output from a completely unpolarized input. This finding is in strong contrast with earlier results that considered static SOI: as it can be proved (using strong, symmetry based arguments), a two-terminal device utilizing any type of (static) SOI can never produce spinpolarized output \cite{KK05}. In the following we show that the physical reason for the temporal spin-polarization we report is the spatial interference of various "Floquet channels" \cite{WC07}, i.e., quantum mechanical plane waves with different wavenumbers.

In the current paper first we analytically determine the Floquet spectrum of electrons in quantum wires with oscillating SOI, and use this result to develop a numerical method for the related time-dependent transport problem. In Sec.~\ref{resultsec}, we present time-resolved results showing that the oscillating SOI generates spin polarization waves.

\begin{figure}[htb]
\centering
\includegraphics[width=8cm]{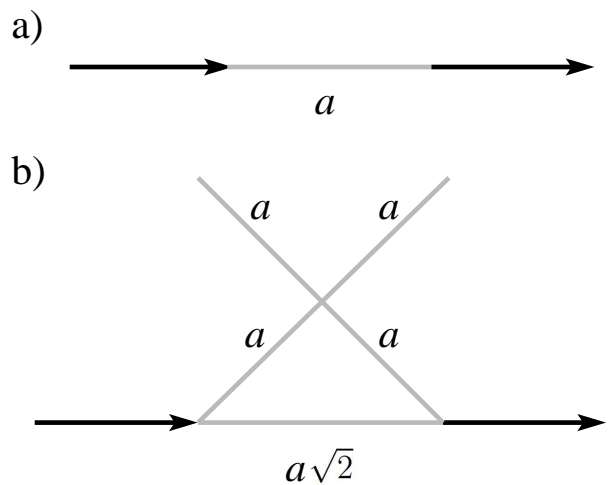}
\caption{The considered geometries. Grey lines correspond to quantum wires with oscillating SOI. We assume no spin-orbit interaction in the input/output leads that are indicated by the black arrows.}
\label{latticefig}
\end{figure}

\section{Model}
\subsection{Oscillating SOI, Floquet quasi-energies}
The building blocks of the devices we consider are straight, narrow quantum wires (see Fig.~\ref{latticefig}). The relevant time-dependent Hamiltonian can be written \cite{MPV04}
as
\begin{equation}
\tilde{H}(t) = \hbar\Omega\left[ \left( -i\frac{\partial}{\partial s}+\frac {\omega(t)}{2\Omega} \mathbf{n}(\mathbf{\sigma}\times\mathbf{e}_z) \right) ^{2}-\frac{\omega(t)^{2}}{4\Omega^{2}}%
\right],
\label{Ham}
\end{equation}
where the unit vector $\mathbf{n}$ points to the chosen positive direction along the wire, and we introduced the characteristic kinetic energy $\hbar\Omega=\hbar^{2}/2m^{\ast}a^{2}$ (with $a$ being the relevant length scale, see Fig.~\ref{latticefig}). The length variable (in units of $a$) along the wire is denoted by $s.$ $\tilde{H}$ depends on time via the strength of the SOI $\omega(t)=\alpha(t)/a$, where it is the Rashba parameter $\alpha$ that can be tuned by external electrodes \cite{NATE97,G00}. In the following we assume:
\begin{equation}
\omega(t)=\omega_{0}+\omega_{1}\cos(\tilde\nu t).
\label{alphat}
\end{equation}
Note that $\tilde{H}$ is linear in $\omega(t)$ [the compact form given by Eq.~(\ref{Ham}) does not show it explicitly, but the quadratic terms cancel each other]. Introducing dimensionless units,  the time-dependent Schr\"odinger equation reads
\begin{equation}
i\frac{\partial}{\partial\tau}|\psi\rangle(\tau)=H(\tau) |\psi\rangle(\tau),
\label{TDSE}
\end{equation}
where $\tau=\Omega t$ and $H=\tilde{H}/\hbar \Omega$. The time-dependent part of the SOI can be written as $\omega_{1}\cos(\nu \tau),$ where $\nu=\tilde\nu/\Omega.$
Note that for $a=100$ nm, $m^{\ast}=0.067m_e$ (GaAs), $\omega_0/\Omega\approx 5$ is in the experimentally achievable range, and $\Omega$ is of the order of $10^{11}$ Hz. For larger samples -- according to the scaling discussed above -- the characteristic frequencies are lower, and therefore the experimentally achievable maximal SOI strength corresponds to lower values of $\omega_0/\Omega.$

Since the Hamiltonian appearing in Eq.~(\ref{TDSE}) is periodic in time, $H(\tau)=H(\tau+T)$ with $T=2\pi/\nu,$ Floquet theory \cite{F883} can be applied. Using a plane wave basis, the spinors \cite{FSP11}
\begin{equation}
\vert\varphi^{\pm}\rangle(s)=\frac{e^{iks}}{\sqrt{2}}\left(\begin{array}{cc}
1 \\
\pm ie^{i\delta},
\end{array} \right)
\label{eigen}
\end{equation}
satisfy
\begin{equation}
H(\tau)\vert\varphi^{\pm}\rangle(s)=\left[k^{2}\pm k \frac{\omega(\tau)}{\Omega}\right]\vert\varphi^{\pm}\rangle(s),
\end{equation}
where $\delta$ is the azimuthal angle corresponding to $\mathbf{n}$ and the wave number $k$ is measured in units of $1/a.$
Since $\left[H(\tau),H(\tau')\right]=0,$ the eigenspinors above can be used to construct solutions to the time-dependent equation (\ref{TDSE}):
\begin{equation}
\vert\varphi^{\pm}\rangle(s,\tau)=\exp\left[-ik^{2}\tau\mp ik\int_0^\tau \frac{\omega(\tau')}{\Omega} d\tau'\right]\vert\varphi^{\pm}\rangle(s).
\label{eig}
\end{equation}
Using the explicit form of the oscillating SOI, and performing the integral in the exponent, we obtain that the two nonequivalent Floquet quasienergies for a fixed wavenumber $k$ are given by
\begin{equation}
\varepsilon^{\pm}(k)=k^{2}\pm k\frac{\omega_0}{\Omega},
\label{quasiE}
\end{equation}
and the "time-dependent basis spinors" read
\begin{equation}
\vert\varphi^{\pm}\rangle(s,\tau)=\exp\left[-i\varepsilon^{\pm}(k)\tau \mp ik \frac{\omega_1}{\Omega\nu} \sin\nu\tau \right] \vert\varphi^{\pm}\rangle(s).
\label{tdbase}
\end{equation}
The frequencies (dimensionless energies) that appear in the exponent for a given $k$ can be seen most directly by applying an appropriate Jacobi-Anger identity \cite{AS65}, leading to
\begin{equation}
\vert\varphi^{\pm}\rangle(s,\tau)=e^{-i\varepsilon^{\pm}(k)\tau} \sum_{n=-\infty}^{+\infty} J_n\left(\frac{k\omega_1}{\Omega\nu}\right) e^{\mp i n\nu\tau} \vert\varphi^{\pm}\rangle(s),
\label{tdbase2}
\end{equation}
where Bessel functions of the first kind \cite{AS65} appear in the expansion.

\subsection{Fitting to the leads}
We assume no SOI in the leads [that are denoted by the horizontal ($x$ direction) black arrows in Fig.~\ref{latticefig}], and consider a monoenergetic input
\begin{equation}
\vert \Psi \rangle_{\mathrm{in}}= e^{i(k_0x-k_0^2\tau)}\vert u\rangle,
\label{input}
\end{equation}
where $\vert u\rangle$ can be an arbitrary spinor.

In order to obtain a time-dependent solution in the whole domain, the spinor valued wave functions have to be joined at the junctions. We require the spinor components to be
continuous. Additional boundary conditions can be obtained by using
the relevant continuity equation
\begin{equation}
\frac{\partial}{\partial \tau}\rho(s,\tau)+
\frac{\partial}{\partial s} J(s,\tau)=0,
\label{continuity}
\end{equation}
where the (unnormalized) electron density is given by
\begin{equation}
\rho(s,\tau)=\langle \Psi \vert \Psi \rangle(s,\tau),
\label{density}
\end{equation}
while the corresponding current density \cite{MPV04} reads
\begin{equation}
J(s,\tau)=2\mathrm{Re}\langle \Psi \vert -i \frac{\partial}{\partial s} +\frac {\omega(t)}{2\Omega} \mathbf{n}(\mathbf{\sigma}\times\mathbf{e}_z) \vert \Psi\rangle(s,\tau).
\label{currentdens}
\end{equation}
$\vert \Psi \rangle(s,\tau)$ above denotes a solution to the  the time-dependent Schr\"odinger equation (\ref{TDSE}) and the inner product and the expectation value appearing in Eqs.~(\ref{density}) and (\ref{currentdens}) are understood in the spinor sense, that is, no spatial integration is involved. As usual, the physical meaning of the continuity equation (\ref{continuity}) is seen most directly by integrating it over a finite spatial domain: The (unnormalized) probability of finding an electron inside the domain changes as a consequence of the currents flowing in/out at the boundaries. Focusing on a certain junction, we obtain that the current that enters the junction, should also leave it -- \emph{always}, i.e., at any time instants. As we shall see, the boundary conditions \cite{G53} described above can be handled conveniently in the frequency domain.

Let $\epsilon_0=k_0^2$ denote the (dimensionless) frequency of the input. A spinor valued wave with this frequency partially enters the domain of oscillating SOI (and partially gets reflected). According to the previous subsection, whenever a frequency component $\epsilon_0$ appears in the solution of the time-dependent Schr\"odinger equation (\ref{TDSE}), an infinite number of additional "Floquet channels" \cite{WC07} corresponding to frequencies
\begin{equation}
\epsilon_n=\epsilon_0 +n\nu
\label{freqs}
\end{equation}
open for transmission (with $n$ being integer). Therefore the frequency components given by Eq.~(\ref{freqs}) should be taken into account. However, these frequencies are also sufficient for the complete description of the problem: Since the frequency resolved fitting equations are linear and generally nondegenerate, they provide a nonzero result only for nonzero input, i.e., for the set of frequencies given by Eq.~(\ref{freqs}).

According to Eq.~(\ref{tdbase2}), inside a domain with oscillating SOI, the relevant frequencies are members of the set (\ref{freqs}), if one of the Floquet quasi-energies (\ref{quasiE}) is equal to $\epsilon_n,$ with an arbitrary integer $n$. The solutions of the equations $\varepsilon^{\pm}(k)=\epsilon_n$ are
\begin{equation}
k^{+}_{1,2}(\epsilon_n)=-\frac{\omega_0}{2\Omega}\pm\sqrt{\frac{\omega^{2}_0}{4\Omega^{2}}+\epsilon_{n}}
\label{kap}
\end{equation}
and
\begin{equation}
k^{-}_{1,2}(\epsilon_n)=\frac{\omega_0}{2\Omega}\pm\sqrt{\frac{\omega^{2}_0}{4\Omega^{2}}+\epsilon_{n}},
\label{kam}
\end{equation}
where the subscripts correspond to the $\pm$ signs in the equations. Using these wave vectors, a general solution that contains all the frequencies relevant for the description of the problem with the monoenergetic input (\ref{input}) can be written as
\begin{equation}
\vert \Psi \rangle_{\mathrm{osc}}=\sum_{n=-\infty}^{+\infty}\sum_{m=1,2}\left(a^+_{nm}|\varphi^{+}\rangle_{nm} + a^-_{nm}|\varphi^{-}\rangle_{nm}\right),
\label{instate}
\end{equation}
where $a^+_{n,m}$ are unknown coefficients, the space and time dependence of the spinors have been omitted, and $|\varphi^{\pm}\rangle_{n,m}$ denote $|\varphi^{\pm}\rangle(s,\tau)$ evaluated at $\varepsilon^{\pm}(k)=\epsilon_n$ and $k=k^{\pm}_{m}(\epsilon_n).$ That is,
\begin{equation}
\vert\varphi^{\pm}\rangle_{nm}=e^{-i\epsilon_n\tau} \sum_{\ell=-\infty}^{+\infty} J_{\ell}\left(\frac{k^{\pm}_{m}(\epsilon_n)\omega_1}{\Omega\nu}\right) e^{\mp i \ell\nu\tau} \vert\varphi^{\pm}\rangle(s).
\label{instate1}
\end{equation}
\bigskip

The relevant spinor valued wave function in the input lead is the sum of $\vert \Psi \rangle_{\mathrm{in}}$ and the reflected part:
\begin{equation}
\vert \Psi \rangle_{\mathrm{ref}}(x,\tau)=
\sum_n  e^{-i(k'_nx+\epsilon_n \tau)} \left( r_n^\uparrow |\uparrow\rangle + r_n^\downarrow  \vert\downarrow\rangle \right),
\label{reflected}
\end{equation}
where
\begin{equation}
k'_n=\sqrt{\epsilon_n},
\label{kprime}
\end{equation}
while $|\uparrow\rangle$ and $|\downarrow\rangle$ can be arbitrary, provided they are orthogonal in the spinor sense. Note that for negative $\epsilon_n,$ $k'_n$ becomes imaginary; by choosing $k'_n=i\sqrt{-\epsilon_n},$ we can ensure that the corresponding waves decay exponentially towards $x=-\infty.$  These "evanescent" solutions can play an important role in our description based on Floquet states. (The wave numbers $k^{\pm}_{i}(\epsilon_n)$ given by Eqs.~(\ref{kap}) and (\ref{kam}) can also be purely imaginary, but in such cases both signs of the square root are allowed, since they contribute to the wave function in a finite domain.)

The transmitted solution in the output lead is analogous to $\vert \Psi \rangle_{\mathrm{ref}},$ only the signs of the wave numbers are opposite due to the different propagation directions:
\begin{equation}
\vert \Psi \rangle_{\mathrm{trans}}(x,\tau)=
\sum_n  e^{i(k'_nx-\epsilon_n \tau)} \left( t_n^\uparrow |\uparrow\rangle + t_n^\downarrow  |\downarrow\rangle \right).
\label{transmitted}
\end{equation}

Eqs.~(\ref{instate}), (\ref{reflected}) and (\ref{transmitted}) describe the spinor valued solutions of the time-dependent Schr\"odinger equation in all spatial domains. In order to take boundary conditions into account, first one has to evaluate these solutions and their spatial derivatives at the junctions (including the internal ones that are not connected to the input/output leads). As one can see easily, Griffith's boundary conditions \cite{G53} mean a system of linear equations for the coefficients appearing in Eqs.~(\ref{instate}), (\ref{reflected}) and (\ref{transmitted}). Although in principle we have an infinite number of equations, since the Bessel functions $J_n$ for a given argument decrease as a function of their index \cite{AS65}, correct numerical solutions can be obtained by limiting ourselves to a finite number of frequencies. The convergence of the Jacobi-Anger expansion as well as the obtained wave functions were carefully checked when calculating the results to be presented in the following.

\section{Results}
\label{resultsec}
\subsection{Oscillation of the spin direction}
The simple straight geometry shown by Fig.~\ref{latticefig}(a) already shows important consequences of the oscillating SOI. Additionally, it can be used to determine the parameter ranges to focus on. Although the term "traverse time" is difficult to interpret when the input is an infinite wave, the ratio of the length $a$ and $c=E(k)/\hbar k$ with a characteristic wave vector $k$ can tell us which SOI oscillation frequency domain is quasistatic. Accordingly, when $\nu\ll k$ (in dimensionless units), SOI oscillation related effects are expected to be weak. Furthermore, by inspecting Eqs.~(\ref{kap}-\ref{instate}), one can see that the oscillating part of the SOI alone does not induce spin precession. However, when $\omega_0\neq 0,$ the wave numbers that belong to different eigenspinor directions are not the same, thus the spin direction related to superpositions have a nontrivial spatial dependence.

\begin{figure}[htb]
\centering
\includegraphics[width=6cm]{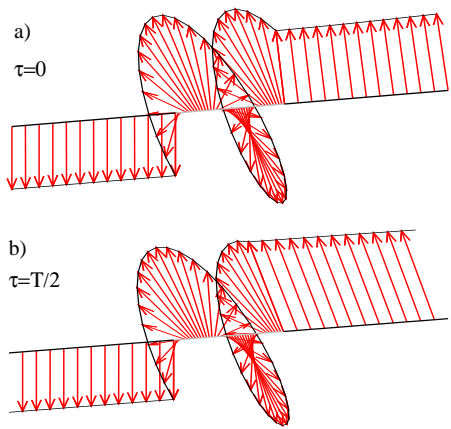}
\caption{Snapshots of the spin direction along a quantum wire. Oscillating SOI is present in the central region (where the color of the wire is gray.) The parameters are $\omega_0/\Omega=\omega_1/\Omega=9,$
 $\nu=1$ and $k_0a=1.5.$ The top panel corresponds to $\tau=0,$ while $\tau=T/2$ for panel b). The thin black line that connects the arrowheads is plotted in order to guide the eyes.}
\label{linespinfig}
\end{figure}

Fig.~\ref{linespinfig} demonstrates this effect. In order to focus on the spin direction alone, the solution $\vert \Psi \rangle(s,\tau)$ (that stands for $\vert \Psi \rangle_{\mathrm{ref}}(x,\tau)+\vert \Psi \rangle_{\mathrm{in}}(x,\tau)$, $\vert \Psi \rangle_{\mathrm{osc}}(s,\tau)$ or $\vert \Psi \rangle_{\mathrm{trans}}(x,\tau)$, depending on the position) has been divided by the space and time dependent electron density given by Eq.~(\ref{density}) (which happens to be nonzero in this case.)
The change of the spin direction along the wire is represented by plotting
\begin{equation}
\mathbf{S}(s,\tau)=\frac{1}{\rho(s,\tau)}\left(\begin{array}{c}
\langle \Psi \vert\sigma_x\vert  \Psi\rangle\\
                  \langle \Psi \vert\sigma_y\vert  \Psi\rangle \\
                  \langle \Psi \vert\sigma_z\vert  \Psi\rangle
                \end{array}\right) (s,\tau),
\label{spindir}
\end{equation}
where the usual Pauli matrices $\sigma_{i}$ appear. More precisely, the arrows shown in Fig.~\ref{linespinfig} point from
$(x,0,0)$ to $(x+S_x, S_y,S_z),$ i.e., they visualize the spin direction in a local coordinate system. The input spinor valued wave function is polarized in the positive $z$ direction in Fig.~\ref{linespinfig}. Note that considering any of the eigenspinors (\ref{eigen}) as input, the spin direction does not change.

Since any difference of the relevant frequencies (\ref{freqs}) is an integer multiple of $\nu,$ the time evolution is periodic ($T=2\pi/\nu$). Fig.~\ref{linespinfig} corresponds to two different time instants, $\tau=0$ (panel a) and $\tau=T/2$ (panel b).  As we can see, the spin direction has a strong spatial dependence in the region, where SOI is present, and when $\omega_0$ is relatively large, there is also a visible time dependence (see around the output lead). %Note, however, that for moderate SOI strengths, it is difficult to identify time dependent effects in the spin direction %for a simple straight wire. (Recall that moderate, experimentally relevant values of $\omega_0/\Omega$ are below $5.$)

\subsection{Generation of propagating density and spin polarization waves}

Although it is difficult to observe in Fig.~\ref{linespinfig}, the oscillating SOI can generate waves that propagate away from the source even in the case of a simple straight wire. The physical reason for the existence of these propagating waves is that SOI oscillations pump energy in the system, populate Floquet states with various frequencies and wave numbers. Boundary conditions "transfer" these populations to regions without SOI, and the interference of these states appear as wave propagation.

\begin{figure}[htb]
\centering
\includegraphics[width=8cm]{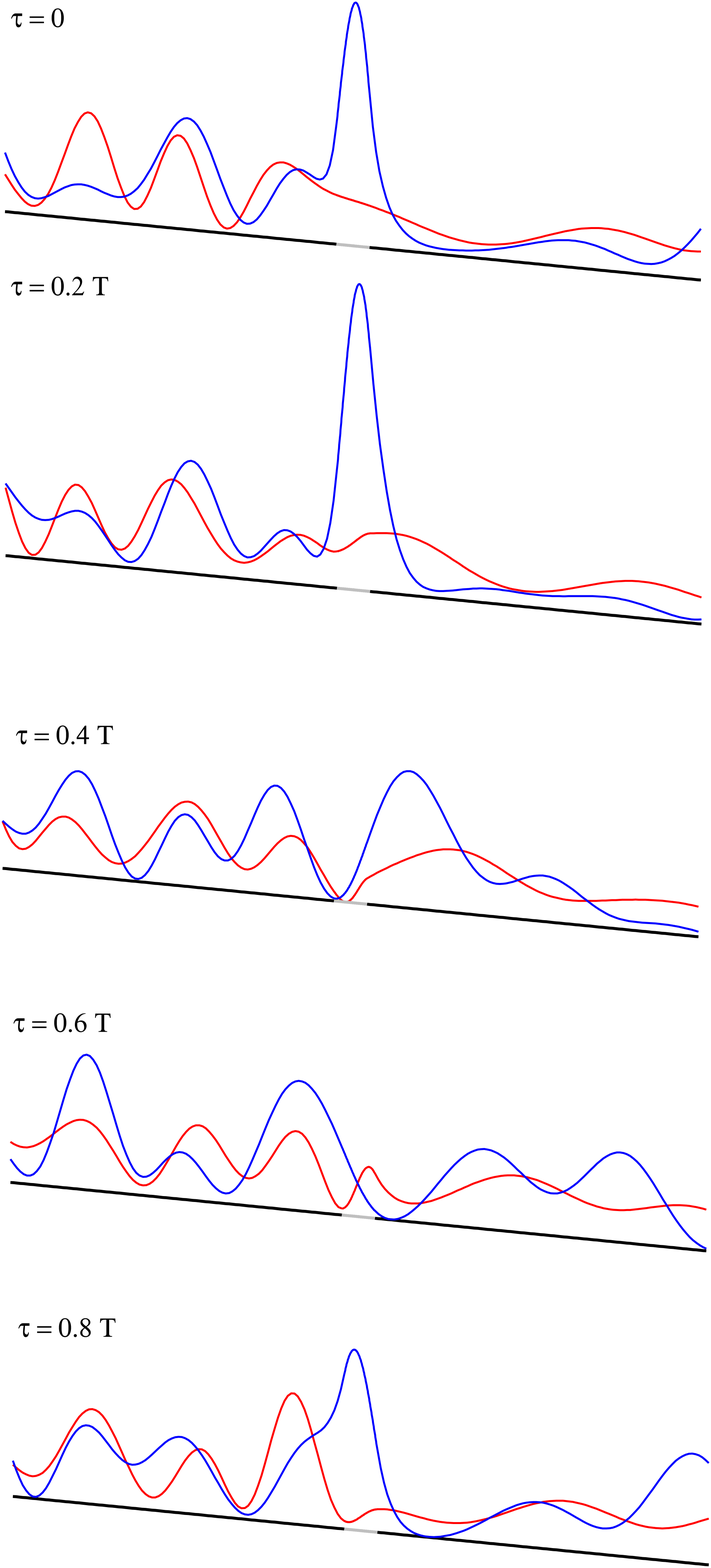}
\caption{Generation of density waves by the oscillating SOI in a quantum wire. The solid red and blue lines show $\rho$ given by Eq.~(\ref{density}) for the $\vert\varphi^{+}\rangle$ and $\vert\varphi^{-}\rangle$ eigenspinor inputs [Eq.~(\ref{eigen})], respectively. The parameters are $\omega_0/\Omega=2.5,$ $\omega_1/\Omega=2.0,$ $\nu=1$ and $k_0a=1.0.$ Time instants when the snapshots were taken are indicated in units of $T=2\pi/\nu.$}
\label{densfig}
\end{figure}

Fig.~\ref{densfig} shows snapshots of the time evolution of the electron density given by Eq.~(\ref{density}). The quantum wire is also shown in this figure, and for each point $(x,y,0)$ of the wire (located in the $z=0$ plane) $\rho$ is plotted as $(x,y,\rho(x,y)),$ see the solid red and blue lines. As we can see, density waves arise and propagate even for moderate SOI strengths.

The figure shows the time evolution of $\rho(s,\tau)$ for both eigenspinors given by Eq.~(\ref{eigen}). Let us recall that the spin direction is conserved for these input spinors, i.e., in contrast to the case shown in Fig.~\ref{linespinfig}, there are no time-dependent spin rotations. On the other hand, however, the space and time dependence of the probability density is different for the two eigenspinor directions. Let us emphasize that this effect is absent for static SOI, when the (time independent) transmission probability is the same for any input spin direction for a two terminal device. This remarkable difference -- on the level of the equations -- can be understood by observing that the wavenumbers (\ref{kap}) and (\ref{kam}) that correspond to the two input spin directions are different when neither the oscillating, nor the static part of the SOI is zero. Consequently, the spatial interference of superpositions of plane waves with these wave numbers produces different patterns for different input eigenspinors. Note that (as we shall see in the following subsection) this fact can lead to temporal spin-polarization -- which is completely absent for the case of static SOI. The related ''no-go theorem'' \cite{KK05} for the equilibrium spin currents is based on symmetry-based considerations, like the unitarity of the scattering matrix, that ensures that the sum of the transmission and reflection probabilities is unity. However, the probability density inside a region with oscillating SOI is generally not constant, thus due to the continuity equation (\ref{continuity}), the magnitude of the current that flows out of the domain does not need to be equal to the current that flows in -- at least not at any time instants. This is the symmetry-related, physical reason for the qualitative difference between the transmission properties of two-terminal devices with static and oscillating SOI.

As a final comment on this point, let us add that although the sum of the reflected and transmitted currents does not need to be always equal to the input current, the periodicity of the problem ensures that the average of these quantities over an oscillation period $T$ satisfies the relation
\begin{equation}
\int_0^T J_{\mathrm{trans}}(\tau) d \tau+\int_0^T J_{\mathrm{ref}}(\tau) d \tau =\int_0^T J_{\mathrm{in}}(\tau) d \tau.
\label{rplust}
\end{equation}
The reflected and transmitted current densities appearing in the integrands are to be calculated using Eq.~(\ref{currentdens}) with $\vert \Psi \rangle=\vert \Psi \rangle_{\mathrm{ref}}$ or $\vert \Psi \rangle=\vert \Psi \rangle_{\mathrm{trans}}$ [see Eqs.~(\ref{reflected}), (\ref{transmitted})] evaluated at the input and output junctions, respectively.
 In practice, we can use these average quantities to check the accuracy of our numerical method: whenever the requirement above is not satisfied within the required numerical precision, more frequency components (\ref{freqs}) have to be taken into account.
\bigskip
\begin{figure}[htb]
\centering
\includegraphics[width=8cm]{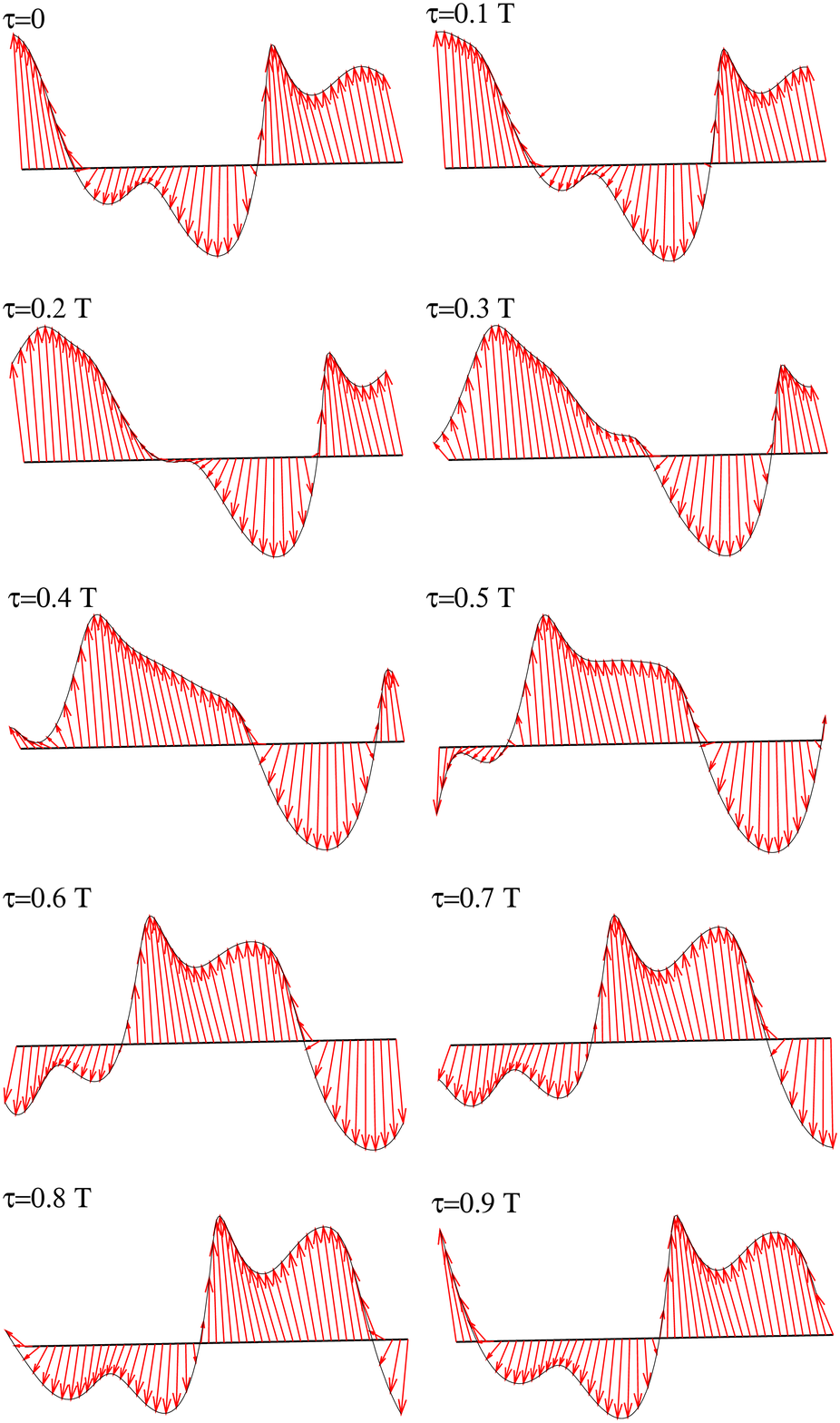}
\caption{Wavelike propagation of the spin direction in the output lead of the loop shown in  Fig.~\ref{latticefig} (b).  The parameters are $\omega_0/\Omega=3.0,$ $\omega_1/\Omega=1.0,$ $\nu=1$ and $k_0a=1.5.$ Time instants when the snapshots were taken are indicated in units of $T=2\pi/\nu.$}
\label{spinloopfig}
\end{figure}

In Fig.~\ref{spinloopfig} we can see the spin direction in the output lead of the geometry shown in Fig.~\ref{latticefig} (b) for several time instants. The input spin is polarized in the positive $z$ direction (not shown), and the spin direction along the lead is visualized in the same way as in Fig.~\ref{linespinfig}. We can clearly identify propagating patterns in this figure.

\subsection{Time-dependent spin polarization}
\begin{figure}[htp]
\centering
\includegraphics[width=8cm]{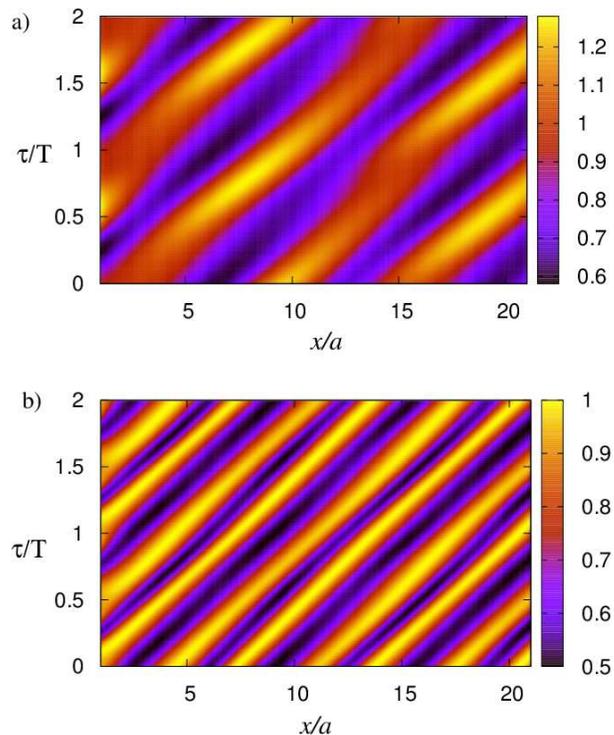}
\caption{The output corresponding to a completely unpolarized input [Eq.~(\ref{Pin})] for the geometry shown in Fig.~\ref{latticefig}(b). $\rho(x,\tau)$ (a) [Eq.~(\ref{density2})] and $p(x,\tau)$ (b) [Eq.~(\ref{purity})] are shown for parameters $\omega_0/\Omega=3.0,$ $\omega_1/\Omega=0.3,$ $\nu=1.0$ and $k_0a=1.0.$ Recall that the density $\rho(x,\tau)$ is not normalized; as a reference, $\rho(x,\tau)=1$ for a single spinor valued plane wave [like the input given by Eq.~(\ref{input})].}
\label{splotfig}
\end{figure}
The results presented so far were obtained using a spin-polarized input. However, according to the previous subsection, interesting results can be expected when the incoming electrons' polarization is random (which is usually the case), their spin state can be described by a quantum mechanical density operator that is proportional to unity:
\begin{equation}
P_{\mathrm{in}}=\frac{1}{2}\left(|\uparrow\rangle\langle\uparrow| + |\downarrow\rangle\langle\downarrow| \right).
\label{Pin}
\end{equation}
Let us note that for a single plane wave, $P$ is constant, but in general it can depend on both space and time. We can calculate the density operator solution of the scattering problem for the input density operator (\ref{Pin}) simply by obtaining $|\Psi\rangle$ for the spin-up and spin-down inputs separately (let us denote these solutions by $|\Psi_\uparrow\rangle$ and $|\Psi_\downarrow\rangle$), and add the corresponding projectors:
\begin{equation}
P(s,\tau)=\frac{1}{2}\left[|\Psi_\uparrow\rangle\langle\Psi_\uparrow|(s,\tau) + |\Psi_\downarrow\rangle\langle\Psi_\downarrow| (s,\tau) \right].
\label{Pop}
\end{equation}
Note that the natural generalization of Eq.~(\ref{density}) in  our case is the following:
\begin{equation}
\rho(s,\tau)=\mathrm{Tr}\left[P(s,\tau) \right].
\label{density2}
\end{equation}

The most interesting result related to $P$ as a function of space and time is that completely pure states can appear in the waves propagating away from the source. As we have already discussed, this effect is in strong contrast with the case of constant SOI, when a loop  rotates the input spin direction -- always in the same way, whatever that direction was \cite{FMBP05,KK05}. The physical reason for the temporal spin-polarization seen in  Fig.~\ref{splotfig} becomes most transparent by recalling Fig.~\ref{densfig}, where the probability densities had different space and time dependent interference patterns for the two eigenspinor inputs. In other words, there are spacetime points where the interference is completely destructive for one of the eigenspinor directions, but not for the other one. This results in a completely polarized spinor. (Note that this is formally analogous to the case reported in Ref.~\cite{KFBP08} for a loop with {\em static} SOI and {\em two} output terminals.)
The results to be presented in this subsection hold for both geometries shown in Fig.~\ref{latticefig}, but -- due to the increased number of paths that can interfere -- the effects are stronger for the triangle loop [Fig.~\ref{latticefig}(b)]. However, let as emphasize that the appearance of the temporal spin-polarization as a physical effect has a weak dependence on the device geometry, e.g., we expect it to be present also for quantum rings (where the technique introduced in the previous section has to be modified) as well as for various polygon geometries \cite{KNV04,KSN06}.

In order to quantify the polarization effects, panel b) of Fig.~\ref{splotfig} shows
\begin{equation}
p(x,\tau)=\frac{1}{\left[\rho(x,\tau)\right]^2}\mathrm{Tr}\left[P^2(x,\tau)\right]
\label{purity}
\end{equation}
as a function of time and the $x$ coordinate in the output lead. Note that the division by the square of the electron density is just for normalization. This function measures the "purity" of the spin state: its range is $\left[\frac{1}{2}, 1\right],$ where the minimum and maximum corresponds to completely unpolarized and 100\% polarized (i.e., pure) states, respectively. According to Fig.~\ref{splotfig}, almost perfect polarization can occur in the output lead for moderate SOI strengths. Additionally, taking a look at panel a) of Fig.~\ref{splotfig}, we can also see that $\rho$ is not zero when the spin is polarized, i.e, there is a finite probability to find the spin-polarized electron at that point.

\begin{figure}[htp]
\centering
\includegraphics[width=8cm]{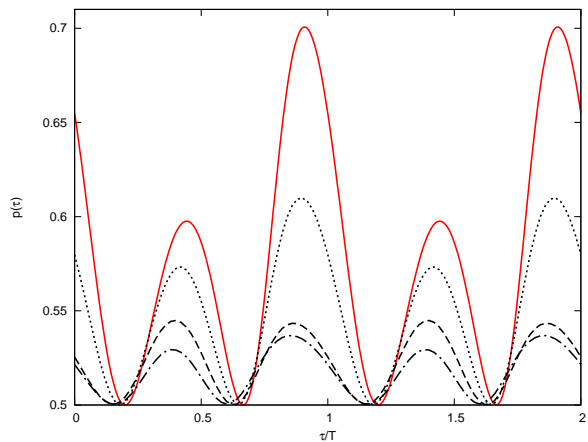}
\caption{The role of spin-dependent scattering mechanisms on the production of spin-polarized currents. $p(\tau)$ [given by Eq.~(\ref{purity})] is shown at $x=3 \ a$  for the linear geometry shown in Fig.~\ref{latticefig}(a). The parameters are $\omega_0/\Omega=5.0,$ $\omega_1/\Omega=0.3,$ $\nu=1.0$ and $k_0 a=1.0.$ The red curve corresponds to the case without scatterers [$D=0$ in Eq.~(\ref{scatt})], while $D=0.6 \ \hbar\Omega$ for the black curves. This relatively strong scattering is present only in the region with oscillation SOI for the dotted curve, only in the output lead for the dashed curve, while the dash-dot line corresponds to the case when there are scatterers in both spatial domains. Technically, we considered three independent, randomly located scattering centers in both regions.}
\label{scattfig}
\end{figure}

\subsection{Disorder related effects}

Finally we investigate the question to what extent our findings are modified by the decrease of the mean free path as a consequence of scattering processes. To this end, similarly to Ref.~\cite{FKP09}, we introduce a random potential
\begin{equation}
U_{scatt}(x)=\sum_n \boldsymbol{U}_n(D) \delta(x-x_n),
\label{scatt}
\end{equation}
where $x_n$ denote uniformly distributed random positions and $\boldsymbol{U}_n(D)$ represents a $2\times 2$ diagonal matrix, with independent random diagonal elements $U_{n1}(D)$ and $U_{n2}(D).$ The argument $D$ is the root-mean-square deviation of the corresponding normal distribution the mean of which is zero. These Dirac delta peaks mean spin-dependent random scatterers and provide an effective model for magnetic impurities of various concentrations: $D=0$ corresponds to the ballistic case, while increasing values mean shorter mean free paths.

In order to see the physical consequences of the scattering processes, we perform a sufficiently large number of computational runs with different realizations of the random potentials (\ref{scatt}) and appropriately average the result (see Ref.~\cite{FKP09} for more details). The polarization effect predicted by our model can be destroyed by scattering induced decoherence in two ways, depending on the position of the scatterers: in the region with oscillating SOI the \emph{generation} of the spin-polarized waves can be hindered, and/or the amplitude of these waves can be decreased by scatterers in the region where they \emph{propagate} (i.e., where there is no SOI). Note that the input spin state is already completely unpolarized, thus there is no need to consider scatterers in the input lead.

As it is shown by Fig.~\ref{scattfig} -- according to the expectations -- the polarization effect gets definitely weaker when we introduce scatterers. Note that $D=0.6\ \hbar \Omega$ that corresponds to this figure means a relatively strong influence on the transport properties, it increases the reflection probability by roughly a factor of two. Moreover, Fig.~\ref{scattfig} visualizes the "worst case", since nonmagnetic scatterers [where $\boldsymbol{U}$ in Eq.~(\ref{purity}) is diagonal (this case is not shown in the figure)] decrease the conductance by a similar amount, but their influence on spin-polarization is considerably weaker. This fact emphasizes the importance of spin coherence length in our findings.

The most interesting fact we can see in Fig.~\ref{scattfig} is that the generation of spin-polarized waves is less sensitive to scattering processes than the propagation of these waves: the same number of scatterers with the same value of $D$ have weaker effect when they are placed in the region where the SOI oscillates. That is, although spin-dependent random scattering decreases the degree of polarization independently from the position of the scatterers, when this process takes place inside the region with oscillating SOI, spin-polarization still can build up, at least partially. This effect can be understood qualitatively: the spin-polarization at the output is stronger when the "interaction region" (where the SOI oscillates and polarization is generated) is longer (with all other parameters being the same). When the spin coherence length decreases below the extension of this region, it defines a new, effective length along which polarization is generated. Thus, realistically, in a sample with long interaction region and disorder, it is the spin coherence length that determines the degree of polarization right at the output, and this is also the length scale that tells us the distance below which the spin-polarized electron waves can be detected in the output lead.

The results of this subsection show that the polarization effect we described earlier in this paper is not extremely sensitive to scattering induced decoherence, thus its experimental observation can be possible.

\section{Summary}
In the current paper we investigated spin-dependent quantum transport through devices in which the spin-orbit interaction (SOI) is time dependent, more precisely, it oscillates. By considering a monoenergetic input, we have shown the emergence of electron density and spin polarization waves propagating away from their source, i.e. the region with oscillating SOI. Additionally, it was demonstrated that simple geometries can produce spin-polarized wave packets even for completely unpolarized input. According to our calculation, this dynamical spin polarization effect appears for realistic, experimentally achievable parameter ranges and remains observable when moderately intensive scattering processes are also taken into account.

In other words, our model suggests a novel source of spin-polarized electrons that can be realized with pure semiconducting materials without the use of external magnetic fields.

\section*{Acknowledgments}
This work was
supported by the Hungarian Scientific Research Fund (OTKA) under
Contract No.~T81364 as well as by the projects
T\'{A}MOP-4.2.2.A-11/1/KONV-2012-0060,
T\'{A}MOP-4.2.2/B-10/1-2010-0012 and
T\'{A}MOP-4.2.2.C-11/1/KONV-2012-0010 supported by the European Union
and co-financed by the European Social Fund.

\bigskip
\providecommand{\newblock}{}

\end{document}